\documentclass[a4paper,12pt]{article}
\usepackage{amsmath}
\usepackage{amsfonts}
\usepackage{accents}
\usepackage{bm}
\usepackage{dsfont}
\usepackage[english]{babel}
\usepackage{graphicx}
\usepackage[usenames]{color}
\usepackage{colortbl}
\usepackage{graphicx}
\usepackage{geometry}
\usepackage{mathtools}
\usepackage{braket}
\usepackage[font=large, labelfont=bf]{caption}
\usepackage{float}
\usepackage{commath}
\usepackage[hidelinks]{hyperref}
\usepackage[font=footnotesize,figurewithin=none]{caption}
\usepackage[caption=false]{subfig}
\usepackage[toc,page,title]{appendix}
\usepackage{comment}

\begin{document}

\centerline {\LARGE{Analytical Investigation of Two-Spin Entanglement}}
\centerline {\LARGE{Generated by Different Types of Bosonic Environments}}
\medskip
\centerline {A. I. Smetana$^1$$^\dagger$, A. R. Kuzmak$^2$$^\dagger$$^{\S}$}
\centerline {\small \it E-Mail: $^1$andriysmetana00@gmail.com, $^2$andrijkuzmak@gmail.com}
\medskip
\centerline {\small \it $^\dagger$Department for Theoretical Physics, Ivan Franko National University of Lviv}
\medskip
\centerline {\small \it 12 Drahomanov St., Lviv, UA-79005, Ukraine}
\medskip
\centerline {\small \it $^{\S}$Haiqu, Inc., 95 Third Street, San Francisco, CA 94103, USA}

\date{\today}
\begin{abstract}
Due to the rapid development of research in the field of quantum physics and quantum information over the past decades, the need to study physical models that can effectively implement quantum computing has increased. An integral part of such models is the environment, which, on the one hand, leads to decoherence in the system, and on the other hand, generates interaction between spins, which in turn allows for the induction of entanglement, which is an integral part of many quantum algorithms. Therefore, it is essential to investigate the impact of the environment on the behavior of quantum systems, enabling the effective implementation of quantum information devices. Here, we consider the time evolution of two spins generated by the interaction through a bosonic environment. The behavior of negativity as a measure of entanglement between spins is studied for different models of environment. As a result, conditions on the parameters of the environment are obtained to achieve the maximum values of entanglement between spins. In this case, environmental models were obtained that minimize the decoherence of the system while maximizing its entanglement. It became possible to derive an effective unitary operator describing the corresponding evolution, since the influence of decoherence was negligibly small.

\end{abstract}

\section{Introduction \label{sec1}}

One of the most important characteristics of quantum systems that is not present in classical physics is entanglement. Quantum entanglement is a phenomenon that appears in multipartite microscopic systems due to interactions between parts of these systems such that the quantum state of each part cannot be described separately, including when particles are separated by a large distance \cite{Einstein1935,Bell1964,Aspect1982}. If a quantum state of the parts of system can be presented as a tensor product of individual quantum states, then it is unentangled, otherwise, it is entangled. This property of nonlocality, which is inherent only to quantum systems, has opened a new direction in information theory, allowing the creation of completely new communication and computing technologies, which are generally called quantum information (for instance, see \cite{Nielsen2010}). 

The implementation of quantum information protocols, including quantum teleportation \cite{Bennett1993,Bouwmeester1997}, quantum cryptography \cite{Ekert1991}, quantum computing \cite{Nielsen2010,Gasparoni2004,Giovannetti2003,Borras2006}, is impossible without entanglement. Various physical platforms have been proposed and experimentally realized for these purposes, such as photons \cite{Bouwmeester1997,Couteau2025}, superconducting qubits \cite{Makhlin2001}, spins of atoms \cite{kane1998,pla2013,Kuzmak2014,Wang2023}, ultracold atoms in optical lattice \cite{Bloch2005,Bloch2013}, tweezer array of atoms \cite{Manetsch2025}, trapped ions \cite{Molmer1999,Bohnet2016,Kwon2024}. A crucial challenge in such systems is the ability to control and readout quantum information while simultaneously protecting the system from environmental interactions that lead to quantum decoherence. In practice, quantum control and measurement are typically implemented using external fields, which inevitably interact with the system and may themselves induce decoherence. Therefore, studying the dynamics of quantum systems under the influence of external fields is of fundamental importance for the development of reliable quantum technologies.

One of the first theoretical studies that provided an analytical description of spin decoherence in a bosonic environment concerned the so-called dephasing model proposed by Luczka in paper \cite{Luczka1990}. This model describes the interaction of the bosonic environment with spin only along one direction. Here, we use this description for a system of two spins placed in an external bosonic environment with different types of spectral density distribution. On the one hand, such an environment leads to decoherence, and on the other hand, it generates interaction between spins, which in turn leads to their entanglement. The study of the influence of the environment on composite quantum systems in the context of the balance between their decoherence and entanglement has been widely studied since the beginning of the 21st century \cite{Yu2003,Dodd2004,Yu2004,Ann2008,Tan2015,Javed2024,Kuzmak2025}.

In paper \cite{Tan2015}, the evolution of two spins that interacts with a common environment described by the Ohmic spectral density was considered. The interaction of spins with the environment was described by the dephasing model \cite{Luczka1990}. The authors studied the time-dependence of the two-spin concurrence as a measure of entanglement that appears due to the interaction between spins through the environment. However, they did not present an analytical expression for concurrence and considered an environment that is described only by the Ohmic spectral density. Unlike previous work, we use negativity to study the dynamics of entanglement of such a system and obtain an exact analytical expression for its behavior depending on the type of environment. In addition to the type of environment described by the Ohmic spectral density \cite{Morozov2012,Chaudhry2013,Ignatyuk2022,Barr2024,Ignatyuk2025}, we investigate entanglement for other types of external environment, such as: single-mode environments and environments described by different Lorentz distributions \cite{Huang2009,Man2014,Zhou2016,Xu2019,Nemati2022,Sun2025}. We also compare the behavior of the entanglement of two spins with spins whose interaction is described by the same effective model, but without the influence of decoherence. The unitary operator in this case describes, with high accuracy, the evolution of two spins with negligible decoherence in the system.

In this paper, we investigate the time evolution of two spins interacting through a bosonic environment (Secs.~\ref{sec2} and \ref{sec3}). The behavior of negativity, used as a measure of entanglement between the spins, is studied for different environmental models (Secs.~\ref{sec4} and \ref{sec5}). The conditions on the environmental parameters required to achieve maximal entanglement between the spins are determined. In Sec.~\ref{sec6}, we propose an effective unitary evolution operator for an idealized model with vanishing decoherence, which accurately reproduces the dynamics in cases where the decoherence effects of the environment are negligibly small.

\section{The model of two spins interacting with an external bosonic environment \label{sec2}}

We consider a system of two spins denoted by "1" and "2" interacting through a common bosonic environment and with an external magnetic field. The expression for the Hamiltonian of the system is written as follows:
\begin{equation}
\begin{aligned}
&H = H_s + H_b + H_{sb}, \\
&H_s= h\left(S^z_1 + S^z_2\right),\quad H_b=\sum_k \omega_k b_k^\dagger b_k,\\
&H_{sb}=\left(S^z_1 + S^z_2\right)\frac{1}{\sqrt{V}} \sum_k (g_k^* b_k + g_k b_k^\dagger),
\label{hamiltonian}
\end{aligned}
\end{equation}
where $S_i^z$ represents the $z$ component of spin operator of the $ith$ spin ($i = 1, 2$), $h$ is the value of the magnetic field, $b_k^\dagger$ and $b_k$ are the creation and annihilation operators of the environment quanta with wave vector ${\bf k}$ and frequency $\omega_k$, $V$ is the volume corresponds to the region where the spin-boson subsystem is located, and $g_k$ characterizes the interaction of spins with bosons. The first, second, and third terms represent the contributions of the system of two spins, the environment, and the interaction between the system and the environment, respectively. Namely, note that the first Hamiltonian $H_s$ describes the interaction of the spins with the magnetic field directed along the $z$ axis. Accordingly, the second Hamiltonian $H_b$ describes the environment as a set of harmonic oscillators, where $\omega_k$ are the frequencies of the harmonic oscillator modes that define the environment. The last Hamiltonian $H_{sb}$ describes the interaction of a system of two spins with a bosonic environment, also called dephasing model \cite{Luczka1990}. We use the system of units, where the Planck and Boltzmann constants are equal to one: $\hbar = 1$, $k_B = 1$. In addition, it is important to emphasize that the spin Hamiltonian $H_s$ commutes with the Hamiltonians $H_b$ and $H_{sb}$: $[H_s, H_b] = [H_s, H_{sb}] = 0$.

We also assume that the spin subsystem and the bosonic environment are initially unentangled and are described by the density matrix
\begin{equation}
\label{initialstate}
    \rho(0) = \rho_s(0) \rho_b(0).
\end{equation}
The initial state of the spin subsystem is the pure state, which can generally be expressed as follows
\begin{equation}
\rho_s(0) =\vert\psi_s(0)\rangle \langle \psi_s(0)\vert,
\end{equation}
where $\vert\psi_s(0)\rangle=\sum_{m_1,m_2=\pm 1} c_{m_1,m_2}\vert m_1, m_2\rangle$ is defined by the complex coefficients $c_{m_1,m_2}$ which satisfy the normalized condition $\sum_{m_1,m_2=\pm 1}\vert c_{m_1,m_2}\vert^2=1$. Here, the set of states $\vert m_1, m_2\rangle$ defines all the projections of two spins on the $z$-axis. The initial state of the bosons is in thermodynamic equilibrium at temperature $T$ and is described by the density matrix
\begin{equation}
\rho_b(0) =e^{-\beta H_b}/{Z_b},
\end{equation}
where $Z_b={\rm Tr}\left[e^{-\beta H_b}\right]$ is the partition function of the environment and $\beta=1/T$.

\section{Evolution of the spins \label{sec3}}

Having started with the state \eqref{initialstate} the quantum evolution of the entire system can be written as follows
\begin{equation}
    \rho(t) = e^{-iHt} \rho(0) e^{iHt}=e^{-i(H_b + H_{sb})t} e^{-iH_s t} \rho_s(0) e^{iH_s t} e^{-\beta H_b} e^{i(H_b + H_{sb})t}/{Z_b}.
    \label{evolutionspe}
\end{equation}
In Appendix~\ref{densitymatrix}, we calculate the evolution of the whole system \eqref{tddensitymatrixfinal}. To obtain the density matrix of the spins, we trace out the bosonic subsystem. After calculations and simplifications, we obtain
\begin{align}
&\rho_s(t)={\rm Tr}_b \rho(t)\nonumber\\
&=\sum_{m_1,m_2=\pm 1}\sum_{n_1,n_2=\pm 1} c_{m_1,m_2}c^*_{n_1,n_2}\vert m_1, m_2\rangle \langle n_1, n_2\vert\exp{\left(-i\frac{ht}{2}(m_1+m_2-n_1-n_2)\right)}\nonumber\\
&\times\exp{\left[-\left(m_1+m_2-n_1-n_2\right)^2\gamma(t)\right]}\nonumber\\
&\times\exp{\left[-i\left(\left(m_1+m_2\right)^2-\left(n_1+n_2\right)^2\right)\Delta(t)\right]}.
\label{spinddensitymatrixfinal}
\end{align}
Here we use that $\left\langle \exp{\left[\gamma b_k^++\alpha b_k\right]} \right\rangle=\exp[\alpha\gamma(\langle b_k^+b_k\rangle+1/2)]$, where $\langle b_k^+b_k\rangle=1/(e^{\beta\omega_k}-1)$. The decoherence coefficients are defined as follows
\begin{equation}
\begin{aligned}
    &\gamma(t) = \sum_k \frac{|g_k|^2}{4V \omega_k^2} (1 - \cos(\omega_k t)) \coth\left(\frac{\beta \omega_k}{2}\right), \\
    &\Delta(t) = \sum_k \frac{|g_k|^2}{4V \omega_k^2} \left(\sin(\omega_k t) - \omega_k t\right),
\end{aligned}
\label{eq:decoherence factors}
\end{equation}
where $\gamma(t)$ describes the loss of coherence, which leads to quantum state mixing, and $\Delta(t)$ leads to the effective Ising interactions between spins.

\section{Entanglement between spins \label{sec4}}

To quantify the entanglement between the spins, we use negativity as a measure \cite{Vidal2002,Plenio2005}. This measure is based on the Peres-Horodecki criterion \cite{Peres1996,Horodecki1996,Horodecki1998} that confirms that subsystem $A$ is entangled with subsystem $B$ if there is at least one negative eigenvalue of their partially transposed density matrix $\rho^{{\rm \Gamma}_{A(B)}}$ with respect to subsystem $A(B)$. This criterion is necessary and sufficient in the case of $2\times 2$ and $2\times 3$ quantum systems. Negativity is defined as the sum of the negative eigenvalues $\Lambda_i$ of $\rho^{{\rm \Gamma}_{A(B)}}$
\begin{eqnarray}
\mathcal{N}(\rho)=\left\vert \sum_{\Lambda_i<0}\Lambda_i\right\vert=\sum_{i}\frac{\vert\Lambda_i\vert-\Lambda_i}{2}.
\label{negativity}
\end{eqnarray}

Since the interaction of the spins with the bosonic bath induces the effective $zz$-Ising interaction defined by the parameter $\Delta(t)$ \eqref{eq:decoherence factors}, the maximal entanglement between spins during the evolution is achieved when the initial state of spins is projected in the $xy$-plane. Therefore, in this section, we consider the behavior of entanglement when the initial state of the spins is projected onto the x-axis and has the form $\vert\psi_s(0)\rangle=1/4\sum_{m_1,m_2=\pm 1} \vert m_1, m_2\rangle$. The detailed derivation of entanglement in the case of the initial state projected along the $x$-axis is presented in Appendix~\ref {derivationnegativity}. The time dependence of the entanglement on the angle between the direction of the effective interaction and the projection of the initial states is presented for different types of environments in Appendix~\ref{deponinitialstate}. Now, using the fact that all parameters of the initial state satisfy $c_{m_1,m_2}=1/4$, the time evolution of the density matrix can be obtained from equation \eqref{spinddensitymatrixfinal}. In the basis $\vert 1, 1\rangle$, $\vert 1,-1\rangle$, $\vert -1,1\rangle$ and $\vert -1,-1\rangle$, it has the form
\begin{equation}
    \rho_s(t) = \frac{1}{4}
    \begin{pmatrix}
    1 & e^{-4\gamma(t)}e^{-4i\Delta(t)} & e^{-4\gamma(t)}e^{-4i\Delta(t)} & e^{-16\gamma(t)} \\
    e^{-4\gamma(t)}e^{4i\Delta(t)} & 1 & 1 & e^{-4\gamma(t)}e^{4i\Delta(t)} \\
    e^{-4\gamma(t)}e^{4i\Delta(t)} & 1 & 1 & e^{-4\gamma(t)}e^{4i\Delta(t)} \\
    e^{-16\gamma(t)} & e^{-4\gamma(t)}e^{-4i\Delta(t)} & e^{-4\gamma(t)}e^{-4i\Delta(t)} & 1
    \end{pmatrix}    
    \label{spinsdensitymatrix}
\end{equation}

Using the definition \eqref{negativity} for the state \eqref{spinsdensitymatrix}, we obtain the negativity between spins (derivation is presented in Appendix~\ref{derivationnegativity}). Analyzing the eigenvalues of the partially transposed density matrix \eqref{eigenvalfornegativity}, we can see that only $\Lambda_2$ is negative. Then the negativity takes the form
\begin{equation}
\begin{aligned}
&\mathcal{N} = \left\vert \sum_{\Lambda_i < 0} \Lambda_i \right\vert = \vert\Lambda_2\vert \\
&= \left|\frac{1}{8}\left(1 - e^{-16\gamma(t)}\right) - \frac{1}{8} \sqrt{\left(1 - e^{-16\gamma(t)}\right)^2 + 16e^{-8\gamma(t)} \sin^2(4\Delta(t))} \right\vert.
\end{aligned}
\label{eq:Negativity_final}
\end{equation}
In the next section, the resulting analytical expression is employed to investigate the behavior of negativity for different types of bosonic baths.

\section{Time dependence of negativity for different spectral densities of boson baths \label{sec5}}

Different types of environments are characterized by their spectral density distribution $J(\omega)$, which in turn determines the behavior of the decoherence parameters $\gamma(t)$ and $\Delta(t)$ \eqref{eq:decoherence factors}. The spectral density includes the information on the distribution over all bath modes of the environment. The rule which allows one to replace the sum over all bath modes with an integral in the decoherence parameters has the form (for example, see \cite{Morozov2012} \cite{Chaudhry2013})
\begin{equation}
    \frac{1}{V}\sum_k|g_k|^{2}f(\omega_k) = \int_{0}^{\infty} J(\omega)f(\omega) \, d\omega.
    \label{eq:sum_to_integral}
\end{equation}
Then the decoherence coefficients \eqref{eq:decoherence factors} take the form
\begin{equation}
\begin{aligned}
    \gamma(t) &= \frac{1}{4}\int_{0}^{\infty}J(\omega)\frac{1 - \cos(\omega t)}{\omega^2}\coth(\beta\omega/2) d\omega, \\
    \Delta(t) &= \frac{1}{4}\int_{0}^{\infty}J(\omega)\frac{\sin(\omega t) - \omega t}{\omega^2} d\omega .
    \label{eq:decoherence factors integral}
\end{aligned}
\end{equation}
Let us consider the behavior of entanglement between spins in the different types of bosonic bath.

\subsection{Single-mode environment \label{subsec5_1}} 

Let us now consider an idealized situation in which all the energy of the environment (boson bath) is concentrated in a single resonance mode $\omega_c$. We can describe this case by using the spectral density in the form of the Dirac delta function:
\begin{equation}
    J(\omega) = \lambda\delta(\omega-\omega_c),
    \label{sddirac}
\end{equation}
where the parameter $\lambda \propto \vert g_k\vert^{2}$ is the strength of the interaction between the spin subsystem and the boson bath. It is worth emphasizing that such a spectral density can idealistically describe the environment inside a resonator, where a standing wave arises at a frequency $\omega_c$, or laser radiation at that frequency. In this case, the decoherence parameters are easily calculated and take the form
\begin{equation}
\begin{aligned}
    \gamma(t) &= \frac{\lambda}{4}\frac{(1-\cos(\omega_ct))}{\omega_c^2}\coth\left (\frac{\beta\omega_c}{2} \right), \\
    \Delta(t) &= \frac{\lambda}{4}\frac{(\sin(\omega_ct) - \omega_ct)}{\omega_c^2}.
\end{aligned}
\end{equation}

\begin{figure}[H]
    \centering
    \includegraphics[width=5.5in, height=4.9in]{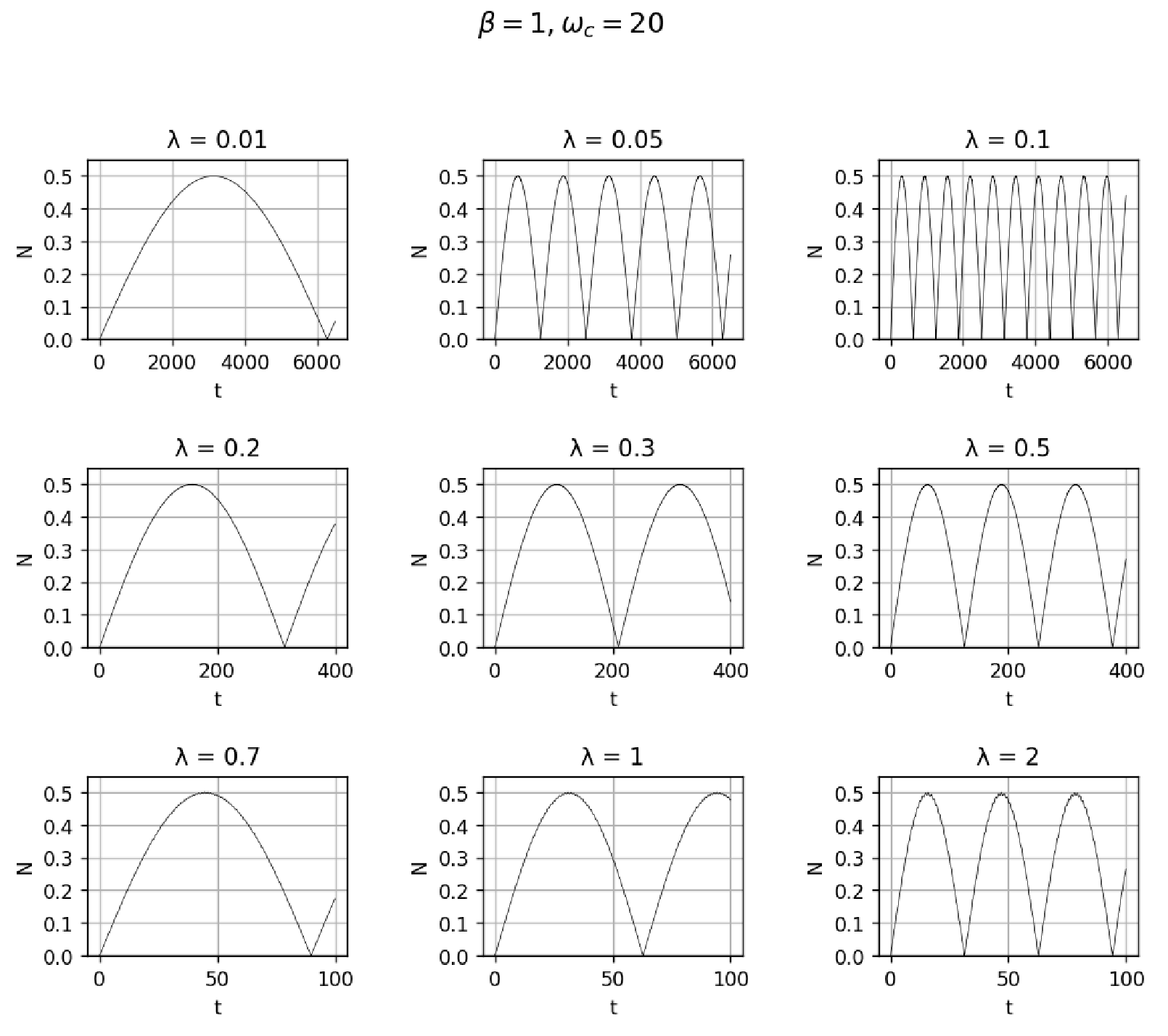}
    \caption{Time-dependence of the negativity at a single-mode spectral density environment with $\beta = 1$, $\omega_c = 20$ and different values of $\lambda$. Increasing $\lambda$ enhances the rate of entanglement dynamics.}
    \label{fig:delta_func1}
\end{figure}

In Fig.~\ref{fig:delta_func1}, we present the time dependence of negativity between spins. As can be seen, depending on the strength of the interaction with the environment $\lambda$, the rate of increase in entanglement is different. The stronger the interaction of spins with the environment, the faster the entanglement grows. This behavior is determined by the fact that $\lambda$ enters the $\Delta(t)$ parameter linearly. This behavior with increasing entanglement inherent in any distribution of the spectral density of the medium. On the other hand, the parameter $\gamma(t)$ leads to damping of the entanglement. For low temperatures in a single-mode environment, the influence of this parameter is negligible. Therefore, in this case, the entanglement always reaches its maximum value. For small coupling strengths ($\lambda = 0.01$ and $\lambda = 0.05$), the growth of the negativity is slow. As the coupling strength increases, the negativity grows faster. For even larger values of $\lambda$, regular oscillations of the negativity emerge. The frequency of these oscillations increases with increasing $\lambda$. Overall, these results demonstrate that the coupling parameter $\lambda$ acts as a key control parameter for the entanglement dynamics.

\begin{figure}[H]
    \centering
    \includegraphics[width=5.5in, height=1.6in]{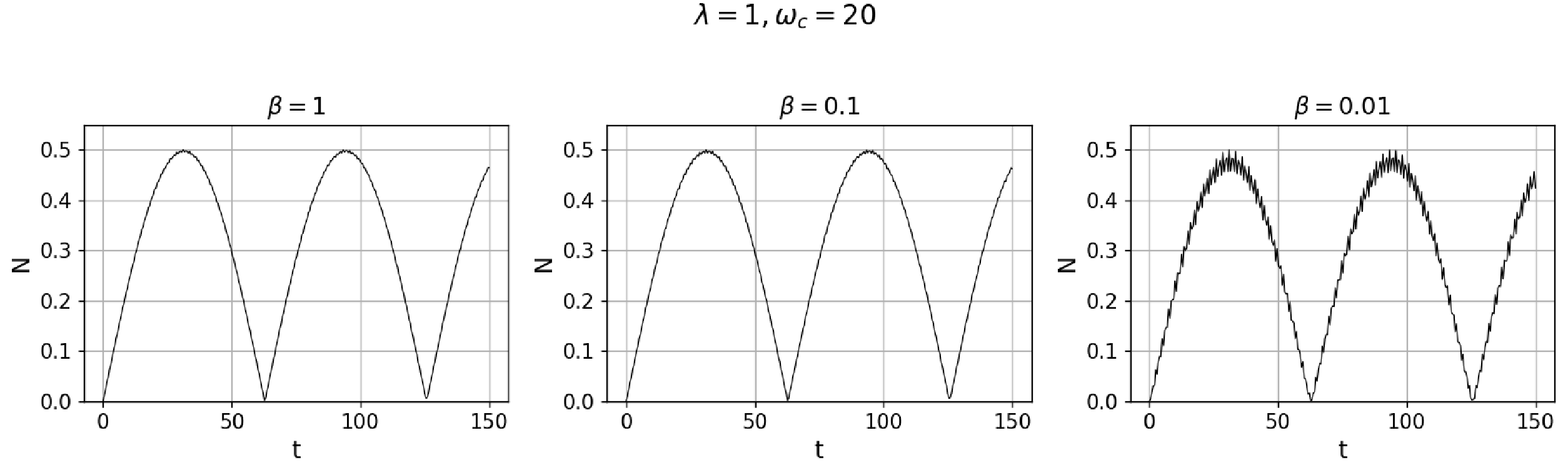}
    \caption{Time dependence of the negativity for a single-mode spectral density environment with $\lambda = 1$, $\omega_c = 20$, and different values of $\beta$. Decreasing $\beta$ preserves the oscillatory entanglement dynamics while introducing additional high-frequency modulations.}
    \label{fig:delta_func2}
\end{figure}

With increasing temperature, the influence of the parameter $\gamma(t)$ becomes more significant, leading to the appearance of short-term oscillations in the time dependence of the entanglement (Fig.~\ref{fig:delta_func2}). As can be seen, when $\beta$ decreases (corresponding to higher temperatures), the oscillatory structure of $\mathcal{N}$ is preserved; however, additional short-term oscillations emerge, which suppress the entanglement due to the enhanced contribution of the parameter $\gamma(t)$. These results demonstrate that, for the Dirac delta spectral density \eqref{sddirac}, the inverse temperature $\beta$ primarily affects the smoothness and stability of the entanglement dynamics rather than the existence of entanglement itself.

\subsection{Ohmic spectral density environment \label{subsec5_2}}

Now we analyze the most well-known dependence for the spectral density used in various spin-boson models (such as \cite{Ignatyuk2022}, \cite{Barr2024})
\begin{equation}
    J(\omega) = \lambda\omega^s\omega_c^{1-s}e^{-\omega/\omega_c},
    \label{eq:Ohmic-like}
\end{equation}
where the parameter $s > 0$ is the so-called ohmicity parameter. Here $\omega_c$ is the cut-off frequency, which means that $J(\omega) \to 0$ at $\omega \to \infty$. The $\omega_c$ defines the frequency range of the boson bath. The value of the parameter $s$ determines different scenarios of interaction between the spin and the environment \cite{deVega2017}. There are the following cases of interactions: the sub-Ohmic case with $0 < s < 1$, the Ohmic case with $s = 1$, and the super-Ohmic case with $s > 1$. It should be noted that this distribution describes various environmental noises. For instance, this distribution describes the spectrum of phonons in solids, and parameters are selected based on experimental data \cite{ramanspectr}. Substituting (\ref{eq:Ohmic-like}) into \eqref{eq:decoherence factors integral}, we obtain the decoherence parameters for the Ohmic environment
\begin{equation}
\begin{aligned}
    \gamma(t) &= \frac{\lambda}{4\omega_c^{-1+s}}\int_{0}^{\infty}(1 - \cos(\omega t))\coth(\beta\omega/2)\omega^{s-2}e^{-\omega/\omega_c} \, d\omega, \\
    \Delta(t) &= \frac{\lambda}{4\omega_c^{-1+s}}\int_{0}^{\infty}(\sin(\omega t) - \omega t)\omega^{s-2}e^{-\omega/\omega_c} \, d\omega.
    \label{eq:decoherence factors integral Ohmnic}
\end{aligned}
\end{equation}
Using these expressions in \eqref{eq:Negativity_final}, we study the behavior of negativity in different regimes of the Ohmnic environment.

\begin{figure}[H]
    \centering
    \includegraphics[width=5.5in, height=4.9in]{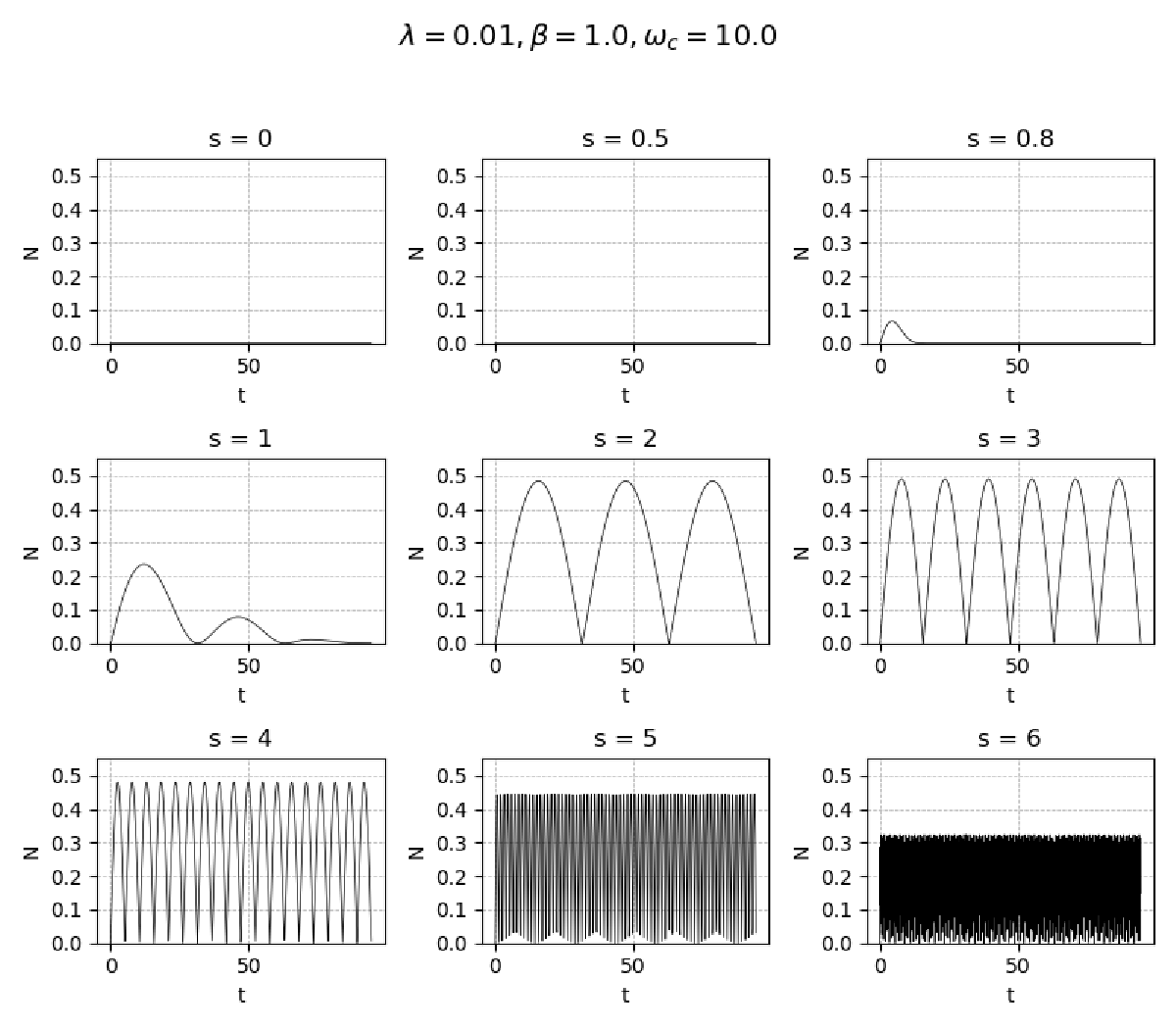}
    \caption{Time dependence of the negativity in an Ohmic-like environment for several values of the parameter $s$. The calculations are performed for $\lambda = 0.01$, $\beta = 1$, and $\omega_c = 10$. One can observe that the behavior of the negativity changes qualitatively with respect to $s$. In the sub-Ohmic regime, the oscillations of $\mathcal{N}(t)$ are strongly suppressed and the entanglement develops rather slowly. Near the Ohmic regime ($s \approx 1$), the oscillatory structure becomes more pronounced, while in the super-Ohmic regime the oscillations acquire larger amplitudes and persist over longer time intervals. In particular, for $s=2$ and $s=4$, the negativity periodically approaches values close to its maximal value.}
    \label{fig:negativity_ohmnicity_all}
\end{figure}

As can be seen in Fig.~\ref{fig:negativity_ohmnicity_all}, the dynamics of entanglement strongly depend on the Ohmicity parameter $s$, which determines whether the bosonic environment belongs to the sub-Ohmic, Ohmic, or super-Ohmic regime. The calculations were performed for a fixed coupling strength $\lambda = 0.01$, inverse temperature $\beta = 1$, and cut-off frequency $\omega_c = 10$. For sub-Ohmic environments, the generated entanglement remains relatively weak, while in the Ohmic case ($s = 1$) the negativity already reaches significantly larger values. A further increase of $s$ into the super-Ohmic regime ($s \in [2,4]$) leads to a considerable enhancement of the entanglement generation, indicating that the spectral structure of the environment plays a crucial role in the formation of quantum correlations between the spins. For $s>4$ the decoherence parameter $\gamma(t)$ begins to dominate, which leads to a decrease in entanglement in the system.

\begin{figure}[H]
    \centering
    \includegraphics[width=5.5in, height=4.9in]{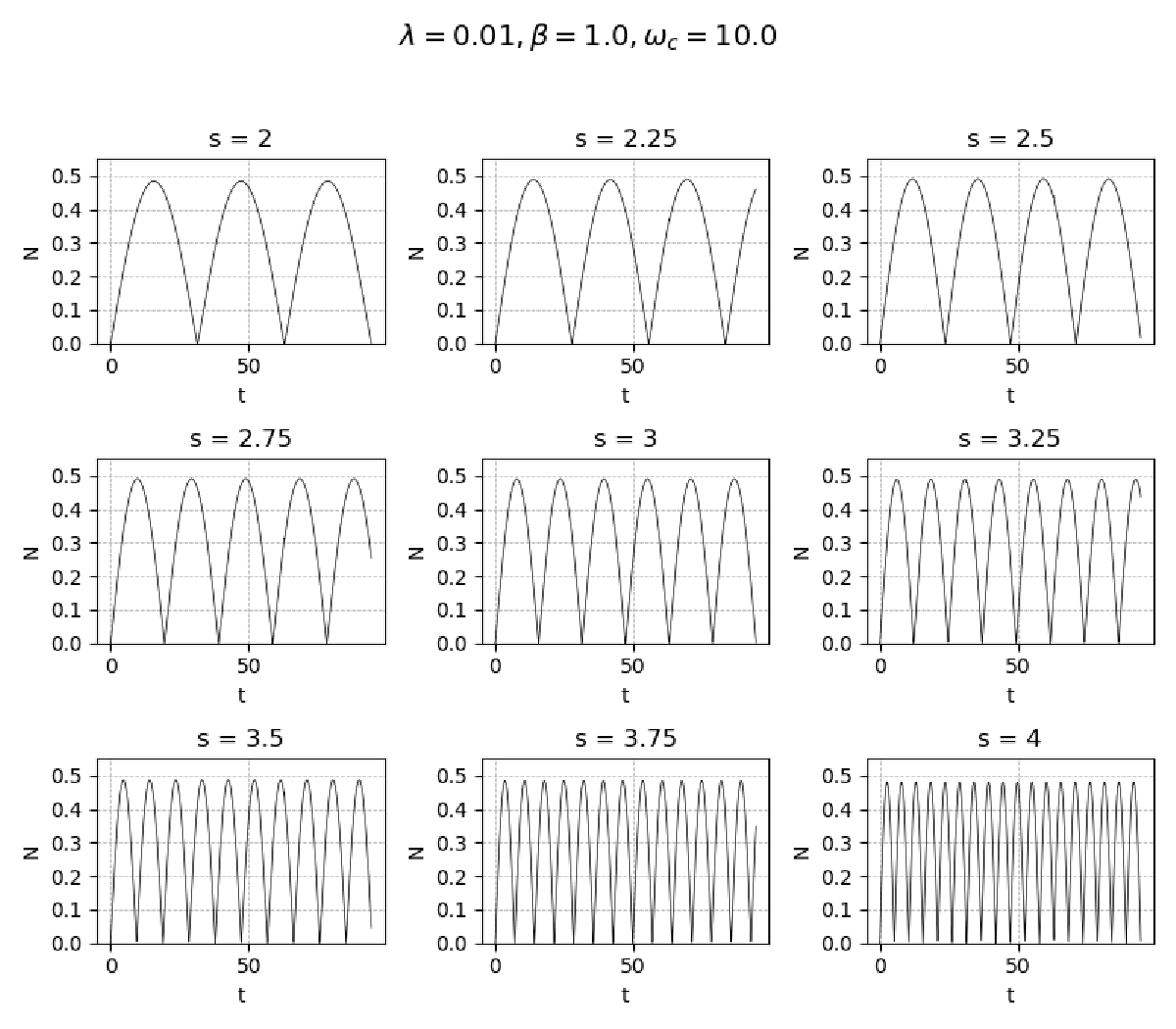}
    \caption{Time-dependencr of the negativity in an Ohmic-like environment for $s\in [2,4]$. This range corresponds to the region where the generated entanglement is maximized. All results are obtained for $\lambda = 0.01$, $\beta = 1$, and $\omega_{c} = 10$.}
    \label{fig:negativity_ohmnicity_best}
\end{figure}

To determine the optimal parameter region more precisely, we performed additional calculations for values of the Ohmicity parameter in the interval $s \in [2,4]$, as shown in Fig.~\ref{fig:negativity_ohmnicity_best}. All calculations were again carried out for $\lambda = 0.01$, $\beta = 1$, and $\omega_c = 10$. It was established that within this interval the negativity reaches its maximal values, corresponding to the most efficient entanglement generation induced by the environment. The interval $s \in [2,4]$ provides the optimal balance between environment-induced spin interactions and decoherence effects, resulting in the strongest and most stable entanglement oscillations. Therefore, engineering the spectral density of the bosonic bath, in particular through the control of the Ohmicity parameter $s$, may provide an efficient mechanism for generating and stabilizing entanglement in open quantum systems.

\subsection{Lorentzian spectral density environment \label{subsec5_3}}

We now consider a bosonic environment characterized by a Lorentzian spectral density. In most general form, the Lorentzian spectral density can be written as
\begin{equation}
J(\omega) = \frac{\lambda}{\pi}
\frac{q\omega^{n}}{(\omega^{2}-\omega_{c}^{2})^{2} + q^{2}\omega^{2}},
\label{eq:lorentz_general}
\end{equation}
where $\omega_c$ defines the resonant frequency of the bath, $q$ is the damping parameter that controls the line width of the spectrum, and $n$ determines the low-frequency scaling of the spectral density, which we take $n=0,1,2$. The Lorentzian spectral density describes an environment with a pronounced resonant peak at $\omega=\omega_c$. Such a spectrum naturally arises in situations where a quantum system interacts with a structured reservoir or a narrow-band radiation field. The parameter $q$ controls the width of the resonance, such that small values of $q$ correspond to a sharply peaked, narrow-band spectrum, while larger 
$q$ lead to a broader distribution. Lorentzian spectral densities model with artificially created environments as a controllable bandwidth (for instance, laser-like sources).
Increasing the power $n$ suppresses low-frequency modes and progressively concentrates the spectral weight around the
resonant frequency $\omega_c$, making the effective spectrum more sharply localized despite an unchanged linewidth parameter $q$. We analyze the behaviour of the entanglement of two spins in the environment with different types of  Lorentzian spectral densities ($n=0,1,2$). 

\begin{figure}[H]
\includegraphics[width=5.5in, height=3.5in]{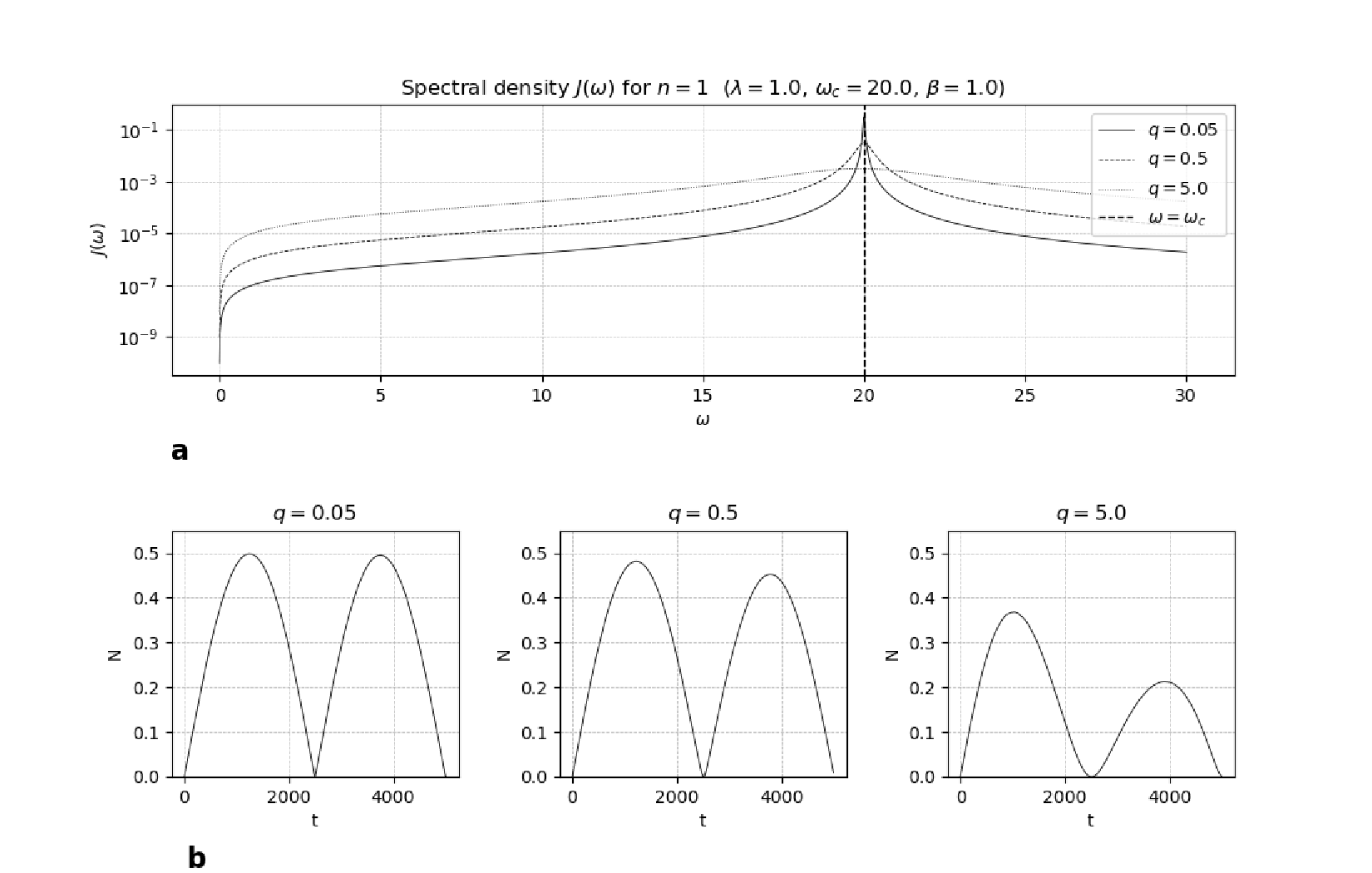}
\caption{Panel (a) shows the Lorentzian spectral density with $n=1$ in \eqref{eq:lorentz_general}, plotted for different values of the damping parameter $q$ at fixed $\lambda = 1$, $\omega_{c}=20$, and $\beta=1$. Panel (b) displays the corresponding time evolution of the negativity $\mathcal{N}(t)$.}
\label{fig:lorentz_omega1}
\end{figure}

We first analyze the case $n=0$. In the case of a Lorentzian spectral distribution with $n=0$, the low-frequency components of the environment play a dominant role. In particular, the decoherence parameter $\gamma(t)$ acquires a strong contribution from the infrared region, leading to what is commonly referred to as an infrared divergence. As a consequence, phase fluctuations accumulate rapidly, and the coherence of the system is suppressed on arbitrarily short time scales. Within this model, decoherence occurs instantaneously, and even in the long-time limit, the system exhibits no recovery of coherence. Since coherence is a necessary resource for entanglement, the persistent and complete loss of coherence prevents the formation of entanglement between the two spins at all times. Therefore, in the presence of Lorentzian noise with $n=0$, the system remains fully unentangled throughout the evolution.

\begin{figure}[H]
\centering \includegraphics[width=5.5in, height=3.5in]{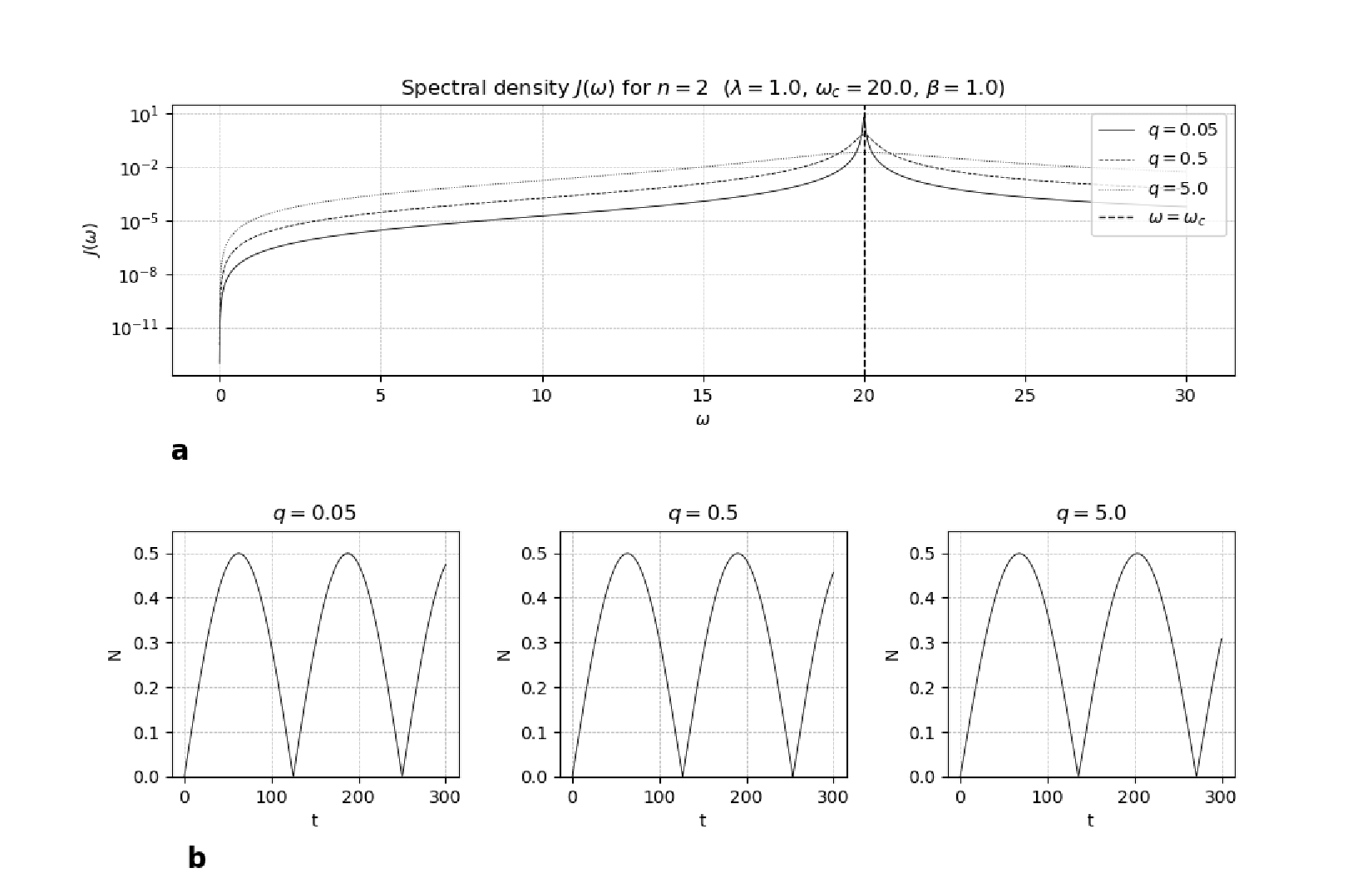}
\caption{Panel (a) shows the Lorentzian spectral density with $n=2$ in \eqref{eq:lorentz_general}, plotted for different values of the damping parameter $q$ at fixed $\lambda = 1$, $\omega_{c}=20$, and $\beta=1$. Panel (b) displays the corresponding time evolution of the negativity $\mathcal{N}(t)$.}
\label{fig:lorentz_omega2}
\end{figure}

For the case $n=1$, increasing the parameter $q$ leads to a significant
suppression of negative (Fig.~\ref{fig:lorentz_omega1}). For small values of $q$ $(q=0.05)$, the
negativity remain almost undamped throughout the  considered time interval. As $q$ increases to $0.5$, a noticeable reduction of the oscillation
amplitude appears, reflecting stronger decoherence effects and a gradual loss
of quantum correlations. For the large value $q=5$, the negativity decays
rapidly and its maximal values are substantially reduced. From a physical point of view, this behavior originates from the broadening of
the spectral density with increasing $q$. A larger linewidth allows the system
to interact with a wider range of environmental frequencies, which increases
the contribution of decoherece.

For the case $n=2$, the overall picture changes significantly (Fig.~\ref{fig:lorentz_omega2}). Despite
variations of the parameter $q$, the amplitude of the negativity remains nearly
unchanged, and the oscillations preserve a stable periodic structure even for
large values of $q$. The principal effect is mainly associated with
modifications of the oscillation frequency and temporal structure, while the
maximal entanglement remains close to $\mathcal{N}=0.5$. This behavior
indicates a substantially weaker influence of decoherence compared with the
$n=1$ case. The factor $\omega^{2}$ enhances the contribution of the $\omega_c$
environmental modes while simultaneously suppressing the other frequency region
of the spectrum. As a result, the system evolves in a more coherent dynamical mode where the
phase evolution dominates over dissipative processes. Therefore, even for large
values of $q$, the entanglement remains robust and exhibits nearly periodic
oscillations without substantial damping.

\section{The negligibly influence of the $\gamma(t)$ parameter on the evolution of two spins \label{sec6}}

Let us analyze the situation where the parameter $\gamma(t)$ is inconveniently small, and the main contribution to evolution is made by the parameter $\Delta(t)$. In such cases, the system almost reaches the maximum entangled states, and the evolution is almost indistinguishable from the evolution if $\gamma(t)=0$. This approximation corresponds to the theoretical idealization of a completely isolated quantum system, where there is no scattering or loss of coherence. This allows us to choose the form of a unitary operator that would approximate such an evolution.

Thus, approximately setting the function $\gamma(t)=0$, the density matrix $\rho_s(t)$ describes the unitary evolution of spins due to the influence of $\Delta(t)$ is as follows
\begin{equation}
    \label{eq:rho_zero_gamma}
    \rho_s(t) = \frac{1}{4}
    \begin{pmatrix}
    1 & e^{-4i\Delta(t)} & e^{-4i\Delta(t)} & 1 \\
    e^{4i\Delta(t)} & 1 & 1 & e^{4i\Delta(t)} \\
    e^{4i\Delta(t)} & 1 & 1 & e^{4i\Delta(t)} \\
    1 & e^{-4i\Delta(t)} & e^{-4i\Delta(t)} & 1
    \end{pmatrix}
\end{equation}
This evolution can be obtained using a unitary operator that describes the time-dependent Ising interactions between spins. Then the density matrix of such evolution takes the form
\begin{align}
&\rho_s(t) = U(t)\vert ++ \rangle \langle ++\vert U(t)^+,\nonumber\\
&U(t) = e^{-i\Delta(t)\left(\sigma_z^{(1)} + \sigma_z^{(2)}\right)^2}.
\end{align}
Note that such an evolution is valid for the case of any initial state $\vert\psi_I\rangle$.
The analytical expression for negativity~\eqref{eq:Negativity_final} in the case of state \eqref{eq:rho_zero_gamma} is as follows
\begin{equation}
    \label{eq:neg_zero_gamma}
    \mathcal{N}(t) = \frac{1}{2}\vert\sin(4\Delta(t))\vert .
\end{equation}

\begin{figure}[H]
    \centering
    \includegraphics[width=5.5in,height=4.9in]{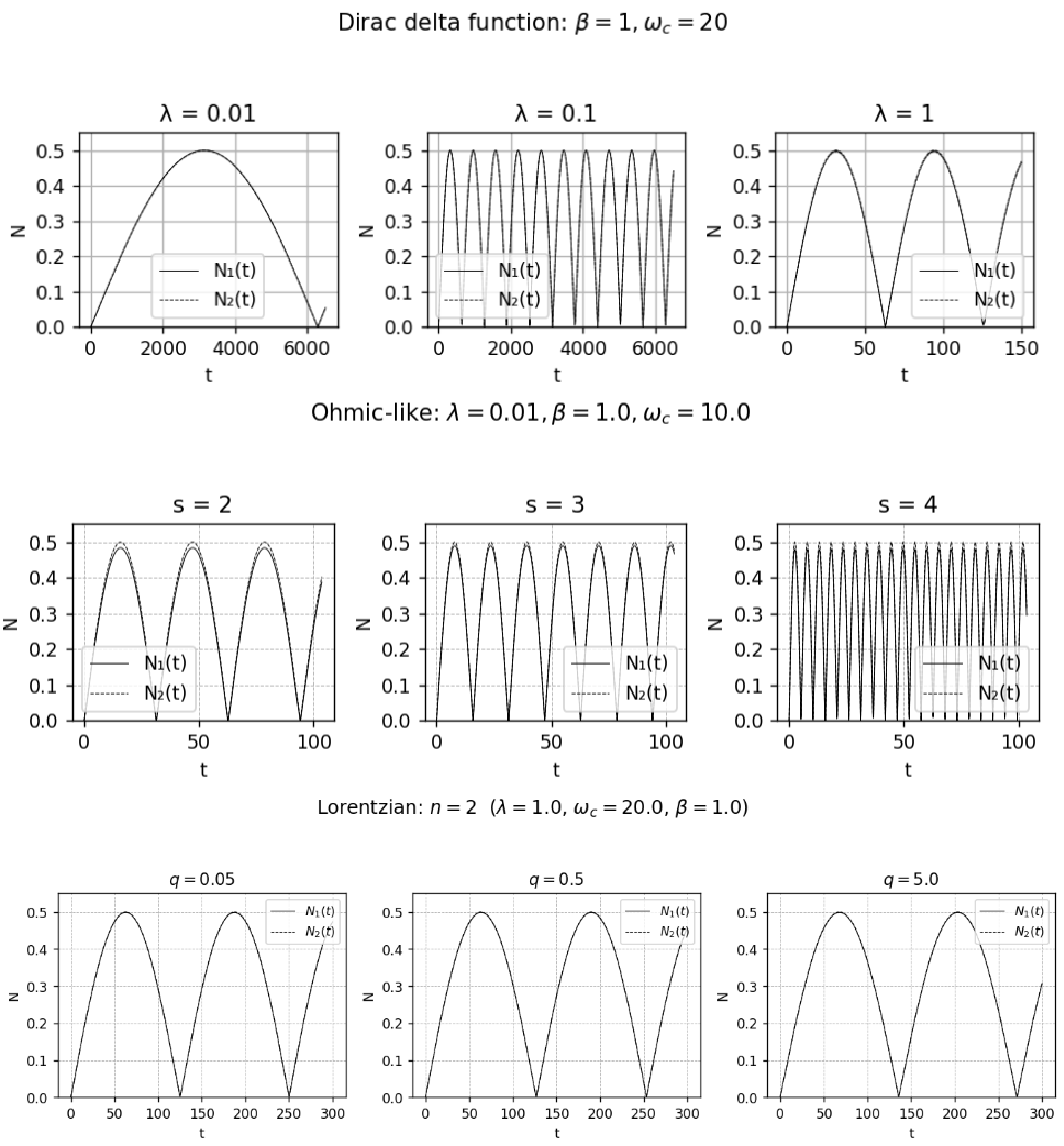}
    \caption{Comparison of the time evolution of negativity for selected parameters at which the influence of the $\gamma(t)$ parameter is negligibly small with
    the idealized time evolution of the negativity \eqref{eq:neg_zero_gamma} for different
    spectral distributions. The upper row corresponds to the Dirac-delta spectral density for $\omega_c=20$ and different coupling strengths $\lambda$, the middle row shows Ohmic-like environments with $\lambda=0.01$, $\omega_c=10$ and Ohmicity parameters $s=2,3,4$, and the lower row presents Lorentzian spectral densities with $n=2$, $\lambda =1$, $\omega_c=20$ different linewidth parameters $q$. All dependencies are presented for low temperature $\beta=1$.}
    \label{fig:zero_gamma_all}
\end{figure}

From the studies in the previous sections, it is easy to see that the cases of environments whose spectral densities are described by the delta function (Subec.~\ref{subsec5_1}), the Ohmic distribution (Subec.~\ref{subsec5_2}) with enough small interaction parameter $\lambda\propto 0.01$ and Ohmicity parameter $s\in[2,4]$, and the Lorentz distribution (Subec.~\ref{subsec5_3}) with a small parameter $q$ for powers $n=1,2$, at low temperatures, are well approximated by the case with $\gamma(t)=0$. This means that in these cases we can achieve maximally entangled states with very high accuracy. In Fig.~\ref{fig:zero_gamma_all}, we compare the dynamics of the negativity obtained for different spectral distributions with parameters that generate small values of $\gamma(t)$ to the idealized case corresponding to $\gamma(t)=0$ described by Eq.~\eqref{eq:neg_zero_gamma}. As can be seen, in all cases, the dynamics of entanglement agrees quite well with the idealized case corresponding to $\gamma(t)=0$.

\section{Conclusions \label{sec7}}

In this work, we have analytically investigated the time evolution of entanglement between two spins coupled to a common bosonic environment within the dephasing model. Using negativity as an entanglement measure, we have derived an exact analytical expression \eqref{eq:Negativity_final} for its time dependence in terms of the decoherence parameters $\gamma(t)$ and $\Delta(t)$ \eqref{eq:decoherence factors}, which encode the effects of the environment through its spectral density distribution. As a result, we have studied three distinct classes of bosonic environments: the single-mode spectral density, the Ohmic family of spectral densities, and the Lorentzian spectral density. For each case, we identify the parameter regimes under which maximal entanglement is generated. This allows us to find the limiting cases where decoherence is negligible ($\gamma(t) \approx 0$). In these cases, we have showed that evolution can be represented with great precision with a unitary Ising-type dynamics governed exclusively by $\Delta(t)$. In this idealized regime, the negativity oscillates periodically without decay, and the system repeatedly returns to maximum entanglement. A compact unitary evolution operator \eqref{eq:rho_zero_gamma} was derived to describe this coherent limit, providing a useful approximation for systems with a specific environmental coupling. From a physical point of view, these results indicate that the bosonic environment cannot be regarded solely as a source of decoherence. Depending on the spectral density parameters, the environment may also play a constructive role by inducing effective correlations between distant quantum subsystems.

\begin{appendices}

\section{Calculation evolution of the spin subsystem \label{densitymatrix}}
\setcounter{equation}{0}
\renewcommand{\theequation}{A.\arabic{equation}}

To calculate the evolution defined by equation \eqref{evolutionspe}, it should be necessary to separate $H_b$ from $\exp{\left(-i(H_b+H_{sb})t\right)}$. Using the Zassenhaus formula \cite{Magnus1954} and its implementation in Mathematica \cite{Cassas2012}, we provide these calculations. For instance, up to the sixth order, this formula is as follows
\begin{align}
&\exp{\left(-i(H_b+H_{sb})t\right)}=\exp{(-iH_bt)}\exp{(-iH_{sb}t)}\exp{\left(\frac{t^2}{2!}[H_b,H_{sb}]\right)}\nonumber\\
&\times\exp{\left(i\frac{t^3}{3!}([H_b,[H_b,H_{sb}]]+2[H_{sb},[H_b,H_{sb}]])\right)}\nonumber\\
&\times\exp{\left( -\frac{t^4}{4!}( [H_b,[H_b,[H_b,H_{sb}]]] + 3[H_{sb},[H_b,[H_b,H_{sb}]]] + 3[H_{sb},[H_{sb},[H_b,H_{sb}]]] )\right)}\nonumber\\
&\times\exp\Bigg(-i\frac{t^5}{5!}([H_b,[H_b,[H_b,[H_b,H_{sb}]]]]+4[H_{sb},[H_b,[H_b,[H_b,H_{sb}]]]] \nonumber\\
& +6[H_{sb},[H_{sb},[H_b,[H_b,H_{sb}]]]]+4[H_{sb},[H_{sb},[H_{sb},[H_b,H_{sb}]]]])\nonumber\\
&+6[[H_b,H_{sb}],[H_b,[H_b,H_{sb}]]]+12[[H_b,H_{sb}],[H_{sb},[H_b,H_{sb}]]] ))\Bigg)\nonumber\\
&\times\exp\Bigg(\frac{t^6}{6!}([H_b,[H_b,[H_b,[H_b,[H_b,H_{sb}]]]]]+5\,[H_{sb},[H_b,[H_b,[H_b,[H_b,H_{sb}]]]]])\nonumber\\
&+10[H_{sb},[H_{sb},[H_b,[H_b,[H_b,H_{sb}]]]]]+10[H_{sb},[H_{sb},[H_{sb},[H_b,[H_b,H_{sb}]]]]]\nonumber\\
&+5[H_{sb},[H_{sb},[H_{sb},[H_{sb},[H_b,H_{sb}]]]]]+10[[H_b,H_{sb}],[H_b,[H_b,[H_b,H_{sb}]]]]\nonumber\\
&+30[[H_b,H_{sb}],[H_{sb},[H_b,[H_b,H_{sb}]]]]+30[[H_b,H_{sb}],[H_{sb},[H_{sb},[H_b,H_{sb}]]]]
\Bigg)\ldots.
\end{align}
The formula of this length is sufficient to discern the pattern and to contract the series under the exponential in subsequent calculations.

Using the explicit form of Hamiltonians $H_b$ and $H_{sb}$ \eqref{hamiltonian}, and taking into account the commutation relations $[b_k,b_{k'}^+]=\delta_{kk'}$, $[b_k,b_k']=[b_k^+,b_{k'}^+]=0$, we obtain the following result
\begin{align}
&\exp{\left(-i(H_b+H_{sb})t\right)}=\prod_k\exp{\left[-i\omega_kb_k^+b_kt\right]}\nonumber\\
&\times\exp{\left[\frac{1}{\omega_k}(1-\cos(\omega_kt))\left(S_1^z+S_2^z\right)\frac{1}{\sqrt{V}}\left(g_k b_k^+-g_k^*b_k\right)\right]}\nonumber\\
&\times\exp{\left[-i\frac{1}{\omega_k}\sin(\omega_kt)\left(S_1^z+S_2^z\right)\frac{1}{\sqrt{V}}\left(g_k b_k^++g_k^*b_k\right)\right]}\nonumber\\
&\times\exp{\left[i\frac{\vert g_k\vert^2}{V\omega_k^2}\left(\omega_kt-2\sin(\omega_kt)+\sin(\omega_kt)\cos(\omega_kt)\right)\left(S_1^z+S_2^z\right)^2\right]}.
\label{sepoperat}
\end{align}
Using the Baker-Campbell-Hausdorff formula, we rearrange the $\exp{\left[-i\omega_kb_k^+b_kt\right]}$ operator to the end of the formula. As a result, we obtain the following expression
\begin{align}
&\exp{\left(-i(H_b+H_{sb})t\right)}\nonumber\\
&=\prod_k\exp{\left[\frac{1}{\omega_k}(1-\cos(\omega_kt))\left(S_1^z+S_2^z\right)\frac{1}{\sqrt{V}}\left(g_ke^{-i\omega_kt} b_k^+-g_k^*e^{i\omega_kt}b_k\right)\right]}\nonumber\\
&\times\exp{\left[-i\frac{1}{\omega_k}\sin(\omega_kt)\left(S_1^z+ S_2^z\right)\frac{1}{\sqrt{V}}\left(g_ke^{-i\omega_kt} b_k^++g_k^*e^{i\omega_kt}b_k\right)\right]}\nonumber\\
&\times\exp{\left[i\frac{\vert g_k\vert^2}{V\omega_k^2}\left(\omega_kt-2\sin(\omega_kt)+\sin(\omega_kt)\cos(\omega_kt)\right)\left(S_1^z+S_2^z\right)^2\right]}\nonumber\\
&\times\exp{\left[-i\omega_kb_k^+b_kt\right]}.
\end{align}
Substituting this expression into the expression for evolution \eqref{evolutionspe}, and once again applying the Baker-Campbell-Hausdorff formula for permutation of the operator $\exp{\left[-\beta\omega_k b_k^+b_k\right]}$, we obtain the time-dependent density matrix of the whole system in the form
\begin{align}
&\rho(t)=\frac{1}{Z_b}\prod_k\exp{\left[\frac{1}{\omega_k}(1-\cos(\omega_kt))\left(S_1^z+S_2^z\right)\frac{1}{\sqrt{V}}\left(g_ke^{-i\omega_kt} b_k^+-g_k^*e^{i\omega_kt}b_k\right)\right]}\nonumber\\
&\times\exp{\left[-i\frac{1}{\omega_k}\sin(\omega_kt)\left(S_1^z+S_2^z\right)\frac{1}{\sqrt{V}}\left(g_ke^{-i\omega_kt} b_k^++g_k^*e^{i\omega_kt}b_k\right)\right]}\nonumber\\
&\times\exp{\left[i\frac{\vert g_k\vert^2}{V\omega_k^2}\left(\omega_kt-2\sin(\omega_kt)+\sin(\omega_kt)\cos(\omega_kt)\right)\left( S_1^z+S_2^z\right)^2\right]}\nonumber\\
&\times e^{-iH_st}\rho_s(0)e^{iH_st}\nonumber\\
&\times\exp{\left[-i\frac{\vert g_k\vert^2}{V\omega_k^2}\left(\omega_kt-2\sin(\omega_kt)+\sin(\omega_kt)\cos(\omega_kt)\right)\left(S_1^z+S_2^z\right)^2\right]}\nonumber\\
&\times\exp{\left[i\frac{1}{\omega_k}\sin(\omega_kt)\left(S_1^z+S_2^z\right)\frac{1}{\sqrt{V}}\left(g_ke^{-(it+\beta)\omega_k} b_k^++g_k^*e^{(it+\beta)\omega_k}b_k\right)\right]}\nonumber\\
&\times\exp{\left[\frac{1}{\omega_k}(1-\cos(\omega_kt))\left(S_1^z+S_2^z\right)\frac{1}{\sqrt{V}}\left(g_ke^{-(it+\beta)\omega_k} b_k^+-g_k^*e^{(it+\beta)\omega_k}b_k\right)\right]}\nonumber\\
&\times \exp{\left[-\beta\omega_k b_k^+b_k\right]}.
\label{tddensitymatrix}
\end{align}
Substituting the explicit form of the density matrix of the spin subsystem, acting on the spin state by the spin part of the operator, and using Weyl's identity to reduce the bosonic operators to a common exponent, we simplify the density matrix \eqref{tddensitymatrix} to the form
\begin{align}
&\rho(t)=\prod_k\sum_{m_1,m_2=\pm 1}\sum_{n_1,n_2=\pm 1} c_{m_1,m_2}c^*_{n_1,n_2}\vert m_1, m_2\rangle \langle n_1, n_2\vert \exp{\left(-i\frac{ht}{2}(m_1+m_2-n_1-n_2)\right)}\nonumber\\
&\times\exp{\left[-i\frac{\vert g_k\vert^2}{4V\omega_k^2}(\sin(\omega_k t)-\omega_k t)\left(\left(\sum_i m_i\right)^2-\left(\sum_in_i\right)^2\right)\right]}\nonumber\\
&\times\exp{\left[-\frac{\vert g_k\vert^2}{2V\omega_k^2}(1-\cos(\omega_k t))\sinh(\beta\omega_k)\sum_i m_i\sum_i n_i\right]}\nonumber\\
&\times\exp\left[\frac{g_k}{2\sqrt{V}\omega_k}	e^{-i\omega_kt}b_k^+\left(1-\cos(\omega_k t)-i\sin(\omega_k t)\right)\left(\sum_i m_i-e^{-\beta\omega_k}\sum_i n_i\right)\right.\nonumber\\
&\left.-\frac{g^*_k}{2\sqrt{V}\omega_k}e^{i\omega_kt}b_k\left(1-\cos(\omega_k t)+i\sin(\omega_k t)\right)\left(\sum_i m_i-e^{\beta\omega_k}\sum_i n_i\right) \right]\nonumber\\
&\times\exp\left[-\beta\omega_kb_k^+b_k\right]/Z_b.
\label{tddensitymatrixfinal}
\end{align}

\section{Derivation of the negativity between spins}
\label{derivationnegativity}
\setcounter{equation}{0}
\renewcommand{\theequation}{B.\arabic{equation}}

Our task is to obtain an analytical formula for negativity between spins defined by the density matrix \eqref{spinsdensitymatrix}. To do this, we need to partially transpose the density matrix. The partially transposed density matrix is as follows
\begin{equation}
\rho^{T_B}(t) = \frac{1}{4}
\begin{pmatrix}
1 & e^{-4\gamma(t)}e^{4i\Delta(t)} & e^{-4\gamma(t)}e^{-4i\Delta(t)} & 1 \\
e^{-4\gamma(t)}e^{-4i\Delta(t)} & 1 & e^{-16\gamma(t)} & e^{-4\gamma(t)}e^{4i\Delta(t)} \\
e^{-4\gamma(t)}e^{4i\Delta(t)} & e^{-16\gamma(t)} & 1 & e^{-4\gamma(t)}e^{-4i\Delta(t)} \\
1 & e^{-4\gamma(t)}e^{-4i\Delta(t)} & e^{-4\gamma(t)}e^{4i\Delta(t)} & 1
\end{pmatrix}
\end{equation}
The eigenvalues $\Lambda$ of this matrix can be found from the equation
\begin{equation}
\left|
\begin{array}{cccc}
\frac{1}{4} - \Lambda & \frac{1}{4}e^{-4\gamma(t)}e^{4i\Delta(t)} & \frac{1}{4}e^{-4\gamma(t)}e^{-4i\Delta(t)} & \frac{1}{4} \\
\frac{1}{4}e^{-4\gamma(t)}e^{-4i\Delta(t)} & \frac{1}{4} - \Lambda & \frac{1}{4}e^{-16\gamma(t)} & \frac{1}{4}e^{-4\gamma(t)}e^{4i\Delta(t)} \\
\frac{1}{4}e^{-4\gamma(t)}e^{4i\Delta(t)} & \frac{1}{4}e^{-16\gamma(t)} & \frac{1}{4} - \Lambda & \frac{1}{4}e^{-4\gamma(t)}e^{-4i\Delta(t)} \\
\frac{1}{4} & \frac{1}{4}e^{-4\gamma(t)}e^{-4i\Delta(t)} & \frac{1}{4}e^{-4\gamma(t)}e^{4i\Delta(t)} & \frac{1}{4} - \Lambda
\end{array}
\right| = 0.
\label{eq:determinant_4}
\end{equation}
This equation is reduced to two quadratic equations
\begin{equation}
\begin{aligned}
    &\frac{1}{4}e^{-8\gamma(t)} \sin^{2}(4\Delta(t)) - \Lambda\left( \frac{1}{4}e^{-16\gamma(t)} - \frac{1}{4} + \Lambda \right) = 0, \\
    &\frac{1}{4}e^{-8\gamma(t)} \cos^{2}(4\Delta(t)) - \left( \frac{1}{2} - \Lambda \right)\left( \frac{1}{4} - \Lambda + \frac{1}{4}e^{-16\gamma(t)} \right) = 0.
\end{aligned}
\end{equation}
Solving these equations, we obtain the eigenvalues
\begin{equation}
\begin{aligned}
\Lambda_1 &= \frac{1}{8}\left(1 - e^{-16\gamma(t)}\right) + \frac{1}{8} \sqrt{\left(1 - e^{-16\gamma(t)}\right)^2 + 16e^{-8\gamma(t)} \sin^2(4\Delta(t))}, \\
\Lambda_2 &= \frac{1}{8}\left(1 - e^{-16\gamma(t)}\right) - \frac{1}{8} \sqrt{\left(1 - e^{-16\gamma(t)}\right)^2 + 16e^{-8\gamma(t)} \sin^2(4\Delta(t))}, \\
\Lambda_3 &= \frac{1}{8}\left(3 + e^{-16\gamma(t)}\right) + \frac{1}{8} \sqrt{\left(3 + e^{-16\gamma(t)}\right)^2 + 16e^{-8\gamma(t)} \cos^2(4\Delta(t)) - 8\left(1 + e^{-16\gamma(t)}\right)}, \\
\Lambda_4 &= \frac{1}{8}\left(3 + e^{-16\gamma(t)}\right) - \frac{1}{8} \sqrt{\left(3 + e^{-16\gamma(t)}\right)^2 + 16e^{-8\gamma(t)} \cos^2(4\Delta(t)) - 8\left(1 + e^{-16\gamma(t)}\right)}.
\end{aligned}
\label{eigenvalfornegativity}
\end{equation}
Only one eigenvalue $\Lambda_2$ can take negative values. Substituting it in the definition of negativity \eqref{negativity}, we obtain equation \eqref{eq:Negativity_final} which determines the entanglement between spins.

\section{Dependence of entanglement on the projection of initial state for different types of environment}
\label{deponinitialstate}
\setcounter{equation}{0}
\renewcommand{\theequation}{C.\arabic{equation}}

A single-qubit pure state in the Bloch-sphere parametrization is defined as
\begin{equation}
|\psi(\theta,\phi)\rangle = \cos\left(\frac{\theta}{2}\right)\vert0\rangle + e^{i\phi}\sin\left(\frac{\theta}{2}\right)\vert 1\rangle,
\label{eq1appc}
\end{equation}
where $\theta$ is the polar angle determining the ratio of amplitudes of the basis states $|0\rangle$ and $|1\rangle$, while $\phi$ is the azimuthal angle that defines their relative phase.  
We now construct a two–qubit product state as the tensor product of two identical
single–qubit states \eqref{eq1appc} with $\phi=0$
\begin{equation}
\begin{aligned}
\lvert \psi\rangle=
\cos^2\left(\frac{\theta}{2}\right)\vert 00\rangle + \cos\left(\frac{\theta}{2}\right)\sin\left(\frac{\theta}{2}\right)\left(\vert 01\rangle+\vert 10\rangle\right)
+ \sin^2\left(\frac{\theta}{2}\right)\vert 00\rangle.
\end{aligned}
\end{equation}
We now consider how the entanglement of a two-spin system evolves in time, assuming that the initial state of each spin is prepared along a direction characterized by the polar angle $\theta$. The analysis is performed for different spectral distributions of the environment. The resulting entanglement dynamics is obtained numerically for a set of polar angles $\theta \in \left\{\frac{\pi} {8},\; \frac{\pi}{4},\; \frac{\pi}{2} \right\}$. For each fixed value of $\theta$, we compute the corresponding negativity $\mathcal{N}(t;\theta)$ of two spins in different bosonic environments.

The numerical results are presented below for three different types of
spectral densities of the bosonic bath: a Dirac delta spectral density, an Ohmic–like spectral density, and the Lorentzian spectral density. Each panel shows the time evolution of the negativity $\mathcal{N}(t;\theta)$ for the set of angles specified above.

\begin{figure}[h!]
    \centering
    \includegraphics[width=5.5in, height=1.5in]{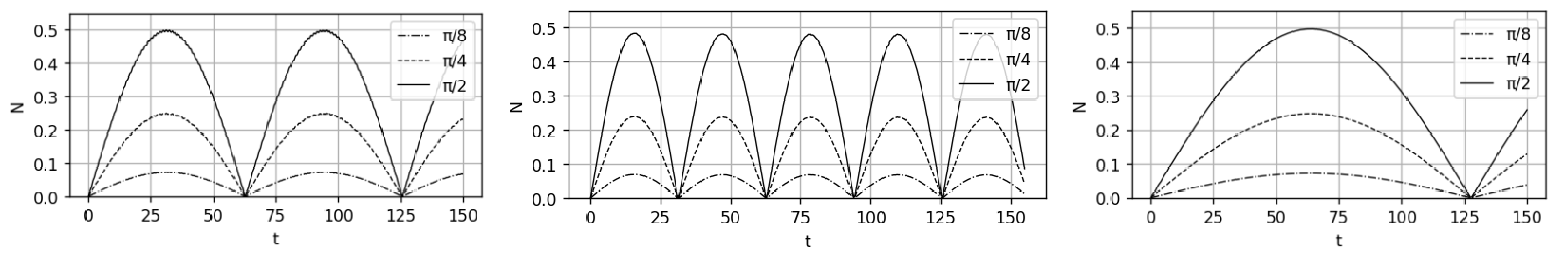}
    \caption {Time-dependence of the negativity for different initial angles $\theta = \pi/8,\ \pi/4,\ \pi/2$ under three distinct environmental spectral densities (from left to right: Dirac delta, Ohmic and Lorentzian spectral densities). In all cases, larger values of $\theta$ lead to systematically higher negativity, highlighting the central role of the initial state in the generation of entanglement.}
    \label{fig:numerical_ohmnic}
\end{figure}

Across all considered environmental spectral densities dependence of the dynamical negativity 
$\mathcal{N}(t;\theta)$ on the initial angle $\theta$ exhibits a clear and universal structure. Since $\theta$ parametrizes the degree of superposition in the initial product state, larger values of $\theta$ correspond to stronger initial coherence and therefore to a greater capacity of 
the system to generate entanglement under environmental dressing. This is reflected in the strict ordering of the negativity curves: for all times $t$, one observes
\begin{equation}
\mathcal{N}(t;\pi/8) \;<\;
\mathcal{N}(t;\pi/4) \;<\;
\mathcal{N}(t;\pi/2),
\end{equation}
independently of the specific form of the spectral density. Physically, this monotonicity highlights the direct mapping between the initial superposition amplitude, encoded in $\theta$, and the system’s ability to build and sustain quantum correlations throughout the evolution.

\end{appendices}

\end{document}